\documentstyle[twoside,fleqn,espcrc2]{article}

\input epsf
\epsfverbosetrue

\title{\vspace*{-28mm}\begin{flushright} {\normalsize NORDITA 98/61 HE/NP} \end{flushright}
Four-quark systems\thanks{Presented by P. Pennanen, 
{\tt Petrus@hip.fi}}}

\author{P.~Pennanen\address{Helsinki Institute of Physics. Currently Nordita, Blegdamsvej 17, 2100 Copenhagen \O, Denmark}, A.M.~Green\address{Dept. of Phys. and Helsinki Inst. of Phys.,
P.O. Box 9, FIN-00014 University of Helsinki, 
Finland}$^\dagger$
and
C.~Michael\address{Theoretical Physics Division, Dept. of Math.
Sciences,
University of Liverpool, Liverpool, UK}\thanks{e-mails: {\tt anthony.green@helsinki.fi,cmi@liv.ac.uk}}}

\begin{document}

\begin{abstract}
In order to understand the binding of four static quarks, flux distributions 
corresponding to these binding energies are
studied in quenched SU(2) and also related to a model for the energies. The potential relevant to string breaking between two heavy-light mesons
is measured in quenched SU(3) using stochastic estimates of light quark propagators with the Sheikholeslami-Wohlert action. 
\end{abstract}

\noindent
\maketitle

Our goal is to understand multi-quark interactions from first principles. 
Since ``multi-quark'' means that the system can be decomposed into more than 
one color singlet, the simplest such system consists of four quarks. 
In this work we try to understand this simple case, relevant for meson-meson 
interactions, hoping that generalization to more complicated cases will
be relatively straightforward. 

We first give a brief introduction to a model that reproduces one hundred 
ground- and excited state energies of four static quarks~\cite{gre:97}. 
These energies have been calculated earlier in SU(2) gauge theory on 
$\approx 20^3 \times 32$ lattices
with $a\approx 0.1$ fm -- a continuum extrapolation has been 
performed~\cite{pen:96b}. The flux distribution producing 
four-quark binding has been measured~\cite{pen:98} and we attempt to relate
it to our model. 

Preliminary results are reported from a more realistic calculation concerning
the potential between two heavy-light mesons in quenched SU(3). The heavy
quarks are taken to be static and the light quark propagators are calculated
using stochastic estimators with an $O(a)$ improved fermion action. This work
also concerns string breaking. 

\vspace{-0.2cm}

\section{The model}

We like to think of the simplest multi-quark system, consisting of four quarks,
as a bridge between a single hadron and interacting multi-hadron systems. 
Earlier, simulations of static four-quark systems have been performed for a 
general set of geometries~\cite[references therein]{pen:96b,glpm:96}. These
quenched SU(2) calculations have produced binding energies up to 120 MeV. 
In order to understand these results a phenomenological model based
on two-body potentials and multi-quark interaction terms has been developed. 

The basis states of the model come from the three possible ways to pair four 
quarks into two mesons in SU(2). These pairing are taken to be in the
ground state ($A,B,C$) or first excited state ($A^*,B^*,C^*$) of the 
two-body potential, giving six
states ($A,\ldots,C^*$) in total. The basic equation of the model is
\begin{equation}
\left[{\bf V}-E(4) {\bf N}\right]=0,
\end{equation}
where ${\bf N}$ contains overlaps of any two states 
($N_{AB} = \langle A|B\rangle, \ldots$), ${\bf V}$ has
the interactions ($V_{AB} = 
\langle A|V|B\rangle , \ldots$) and $E(4)$ is the four-quark energy. 
The binding energy is defined as $B(4)=E(4)-\min(V_{AA},V_{BB},V_{CC})$, i.e.
as the difference of the four-quark energy and the energy of the lowest-lying
two-body pairing. 

A central element in the model is a multi-quark interaction term $f$, defined
as $\langle A|B\rangle =\langle B|C\rangle =\langle A|C\rangle =-f/2$. For 
small distances $f=1$ and for large $f=0$. The potentials $V$ have
a perturbative one-gluon exchange prefactor and lattice two-body potentials 
$v_{ij}$, giving the form $V_{ij}=-\frac{1}{3}\sum_{i\le j} \vec{\tau}_i \cdot
\vec{\tau}_j v_{ij}$. This expression alone would lead to unphysical van der 
Waals forces, which are removed by the $f$. The interaction between
basis states $A$ and $B$ is thus $V_{AB}=-\frac{f}{2}(V_{AA} +V_{BB} - 
V_{CC} )$.

The factor $f$ can be parameterized as $f=\exp(-b_s k_f S)$, where
$b_s$ is the string tension and $S$ the minimum area bounded by 
the four quarks. A simple version of the model with just states $A,B$ and 
$f=1$ is reproduced by perturbation theory to $O(g^4)$\cite{lan:95}. 

For quarks at the corners of a regular tetrahedron binding increases
with size, which suggests two-body excitations are relevant, 
as they are closer to the ground state for larger distances. Therefore,
we have included excited basis states in our model. This gives new $f$-factors
$\langle A^*|B^*\rangle  = \ldots =-f^c/2, \ \langle A^*|B\rangle  = 
\ldots =-f^a/2$, the former of which is parametrized like $f$ and the latter
as $f^a=b_S f^a_S S\exp(-b_s k_a S)$. The interaction
matrix elements involving two-body excitations are parameterized in a 
manner similar to the ground state matrix elements, e.g. 
$\langle A^*|V|B^*\rangle  = -\frac{ f^c}{2}\left[V_{AA}+V_{BB}-V_{CC}+
c_0(V^*_{AA}+V^*_{BB}-V^*_{CC})\right]$.

When this model is fitted to 100 ground and excited state four-quark energies
it turns out that $f_c$ is always consistent with one, suggesting that 
two-body excitations interact in a perturbative manner. We are left
with four independent parameters 
$k_f= 1.5(1),\ k_a=0.55(3), \ f^a_S=0.51(2),\ c_0=3.9(3)$ that
fit the energies with $\chi^2/{\rm d.o.f.}\approx 1$. The values of these parameters
are stable when energies from smaller lattice spacings are used. If
the excited basis states are left out we get $\chi^2 \approx 3.2$, which
mostly comes from the maximally degenerate regular tetrahedra.

\vspace{-0.2cm}

\section{Four-quark flux distribution}
 
The flux distribution is measured in quenched SU(2) using 
$$
 f_R^{\mu \nu}({\bf r})=\left[{\langle
W(R,T)
  \Box^{\mu \nu}_{\bf r}\rangle}
-\langle W(R,T)\rangle \langle \Box^{\mu \nu}\rangle
\over {\langle W(R,T)\rangle} \right],
$$
where $W(R,T)$ is the Wilson loop, or more generally, a combination of
Wilson loops corresponding to the ground or excited state of the multi-quark
system. Since extracting a signal is not easy we use link-integration on the 
Wilson loops. As in the calculation of energies, we utilize a variational
approach in order to extract the ground or an excited state of the system.
The variational basis is formed of Wilson loops with various levels of fuzzing.

We can check the accuracy
of our flux measurement with lattice sum rules. These sum rules relate
spatial sums over fields to energies via $\beta$-functions such as
$b\equiv \partial g/\partial \ln a = 2(S+U)$, e.g.
$$
E + E_0  =  \sum  [S ({\cal E}_x+{\cal E}_y+{\cal E}_z) +
              U ({\cal B}_{z}+{\cal B}_{y}+{\cal B}_{x})]. \\
$$
Using our earlier two-quark results~\cite{pen:97b} for $S,U$ this relation
allows us to compare the sum over the distribution to a much more accurately
measured energy. This is especially convenient when the binding is considered;
$E_0$ is a lattice artefact from quark self-energies, 
and is removed for the binding energy. 

Three-dimensional pictures of our results (Fig.~1) can be viewed at 
{\tt www.physics.helsinki.fi/\~{}ppennane/pics/}. The binding distribution
is found to have approximately constant height in between the quarks, whereas
the first excited state has a ``cloverleaf'' symmetry with sign changes.
Replacing energies with flux distributions in our model without excitated
basis states gives $k_f<1$,
which agrees with fits to energies~\cite{pen:98}. When excited basis states
are included the ground-state interaction has shorter range; $k_f >1$, 
suggesting that ground state potentials are important at small distances
and excited ones at larger distances. 

\begin{figure}[h]
\vspace{-0.3cm}
\hspace{0cm}\epsfxsize=110pt\epsfbox{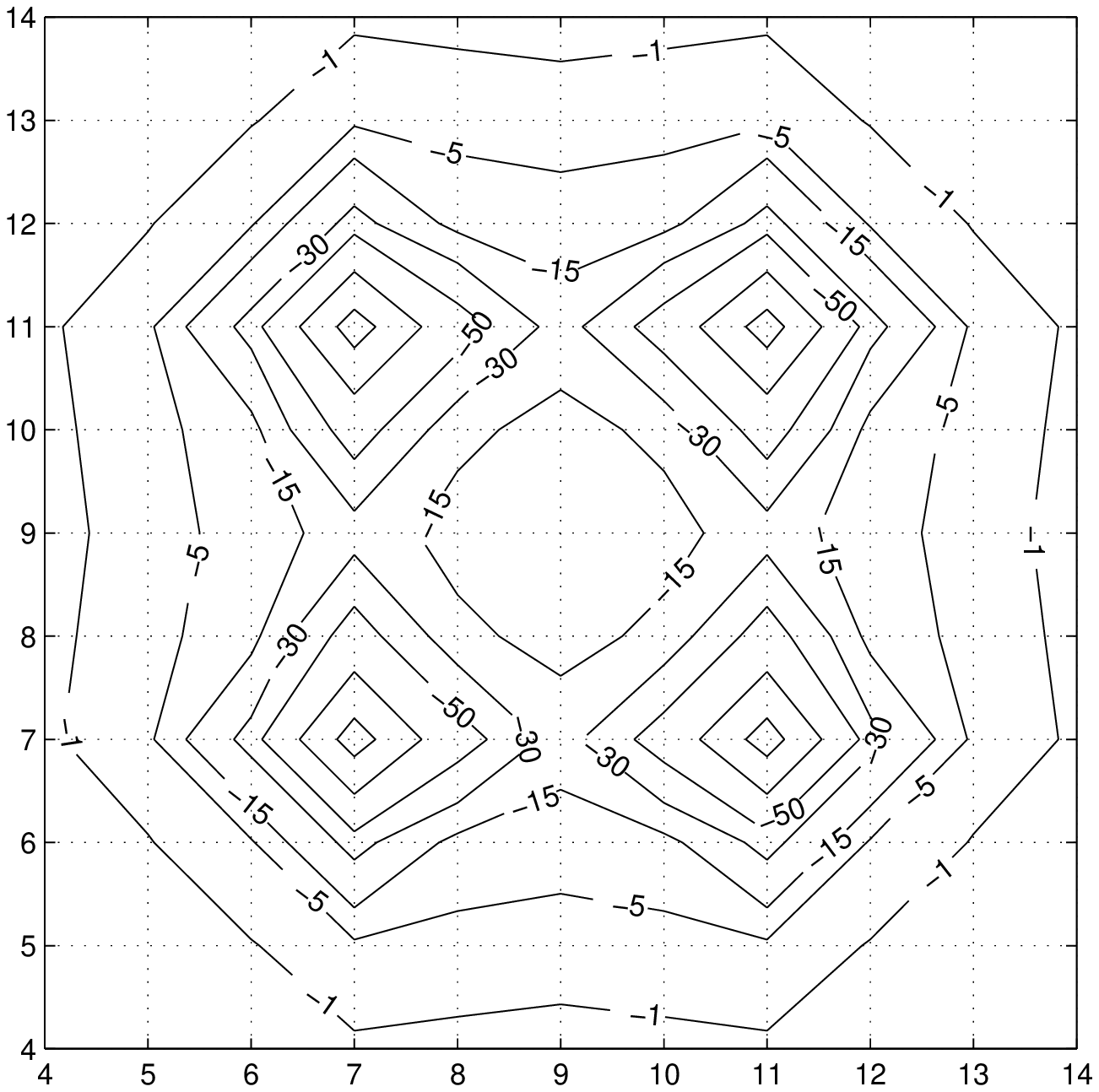}\epsfxsize=110pt\epsfbox{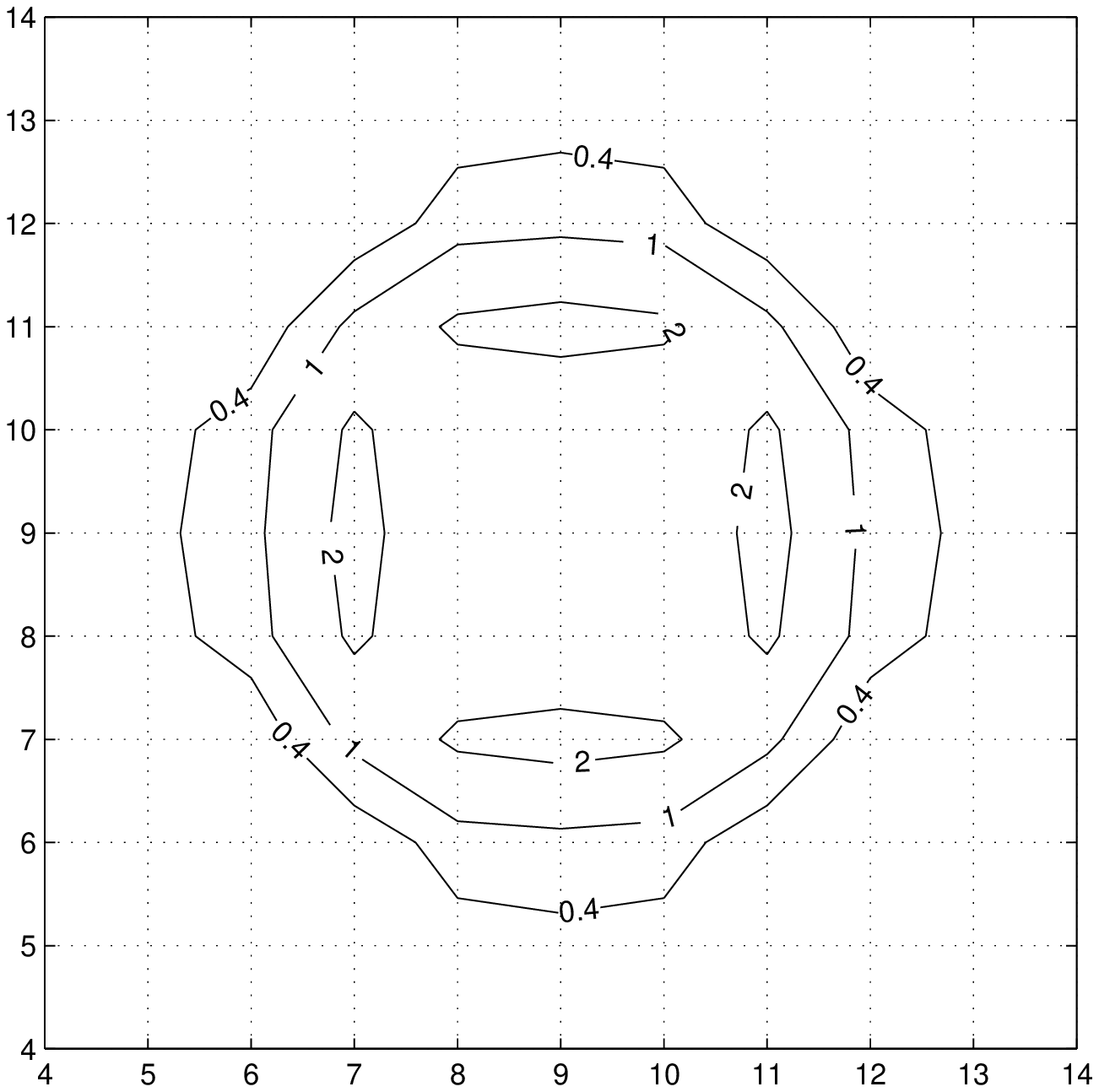} \\
\hspace{0cm}\epsfxsize=110pt\epsfbox{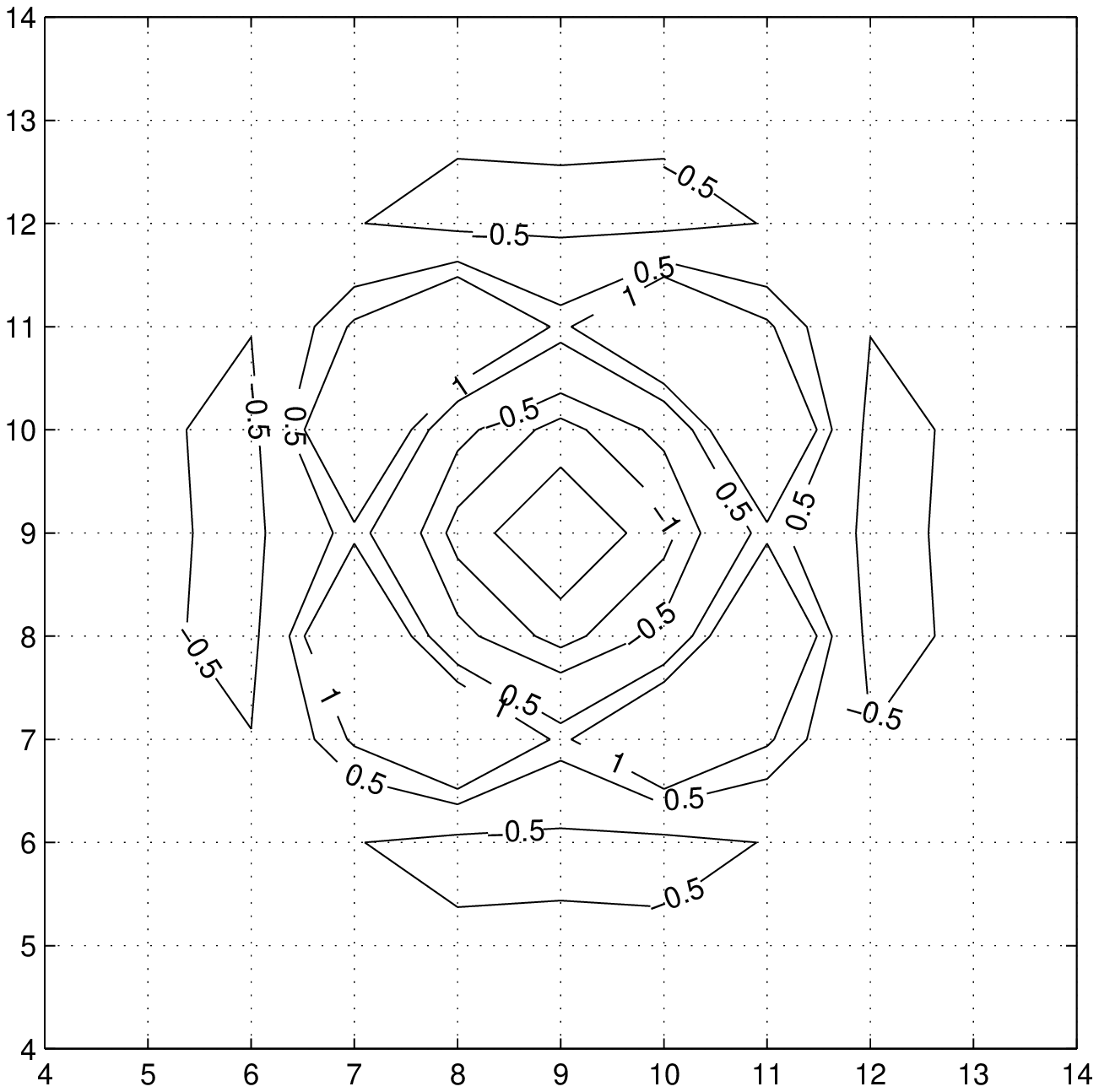}
\vspace{-0.7cm}
\caption{The distribution of energy around four static quarks in a square with side $R=4$ corresponding to a) 
total energy b) binding energy 
c) first excited state binding energy. } 
\label{fflux}
\vspace{-0.5cm}
\end{figure}


\section{Two heavy-light mesons}

We calculate the potential between two heavy-light mesons as a function of the 
heavy quark separation $R$ in quenched SU(3) at $\beta=5.7$. The heavy quarks 
are taken to be static and the light ones have approximately the $s$ 
mass ($M_{\rm PS}/M_{\rm V}=0.65$) with the propagators obtained as 
stochastic pseudofermionic 
estimates with maximal variance reduction~\cite{mic:98}. The principal 
advantage of the stochastic estimates is the ability to get propagators
to and from most points on the lattice; hence the nickname ``all-to-all''. 

\begin{figure}[h]
\vspace{-0.7cm}
\begin{center}
\epsfxsize=230pt\epsfbox{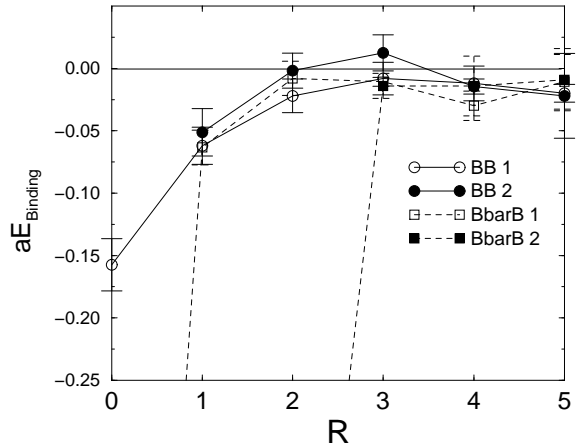}
\end{center}
\vspace{-1.8cm}
\caption{Preliminary results for potential between two $B$ mesons or a $B$ and $\bar{B}$ -- the data
marked with a ``2'' includes light-quark exchange. Here $aM_B=0.863(1)$.}
\label{fb4}
\vspace{-0.3cm}
\end{figure}

At $R=0$ the meson-antimeson ($B\bar{B}$) system should have a mass 
similar to a pion, 
while the $BB$ system should resemble a heavy-light baryon; the 
former is indeed observed. The preliminary results in Fig.~\ref{fb4} do not 
show the one-$\pi$ exchange effect of deuson models which leads
to differences in the $BB$
and $B\bar{B}$ potentials at large distances. As 
expected, at small distances the $B\bar{B}$ potential is more attractive.

As discussed in other papers in these proceedings~\cite{pro}, the heavy-light 
meson-antimeson system is relevant to string breaking. 
Unfortunately 
calculations of this system at large enough distances are very hard, but
the all-to-all propagator scheme offers a way to overcome the difficulties.
In our case the $B\bar{B}$ and static $Q\bar{Q}$ energies become equal at
$R\approx 1.3$ fm~\cite{mic:98b}; this is reached by our ongoing calculations 
on a $16^3\times 24$ lattice. A variational operator mixing 
analysis~\cite{mic:92}
similar to that presented by Philipsen in a Higgs model is then performed.   
By running our code on UKQCD unquenched configurations we hope to shed light
on the differences due to sea quarks in the mixing region.

{\bf Acknowledgement:}
P.P. thanks the Finnish Academy and E. and G. Ehrnrooth Foundation for 
funding.

\vspace{-0.2cm}

\newcommand{\href}[2]{#2}\begingroup\raggedright\endgroup

\end{document}